\documentclass[twocolumn]{aastex631}
\graphicspath{{./}{Figures}}

\def\CII  {[C{\small{II}}]}
\def\micron {$\mu$m}

\shorttitle{Shocked [CII] emission in Arp 25}
\shortauthors{Fadda, Sutter, Minchin, Polles}

\begin{document}

\title{Shock enhanced [CII] emission from the infalling galaxy Arp~25~\footnote{Data obtained with FIFI-LS and HAWC+ onboard SOFIA}}

\correspondingauthor{Dario Fadda}
\email{darioflute@gmail.com, dfadda@stsci.edu}

\author[0000-0002-3698-7076]{Dario Fadda}
\affiliation{
Space Telescope Science Institute, 3700 San Martin Drive, Baltimore, MD 21218, USA}
\author[0000-0002-9183-8102]{Jessica S. Sutter}
\affiliation{University of California San Diego, 9500 Gilman Drive, La Jolla, CA 92093, USA}
\author[0000-0002-1261-6641]{Robert Minchin}
\affiliation{
NRAO Array Operations Center, P.O. Box O, 1003 Lopezville Road, Socorro, NM 87801-0387, USA}
\author[0000-0003-0347-3201]{Fiorella Polles}
\affiliation{SOFIA Science Center, USRA, NASA Ames Research Center, M.S. N232-12 Moffett Field, CA 94035, USA}



\begin{abstract}
We present SOFIA observations with HAWC+ and FIFI-LS of the peculiar galaxy Arp~25, also known as NGC~2276 or UGC~3740, whose morphology is deformed by its impact with the intra-group medium of the NGC~2300 galaxy group. These observations show the first direct proof of the enhancement of [CII] emission due to shocks caused by ram pressure in a group of galaxies. By comparing the [CII] emission to UV attenuation, dust emission, PAH, and CO emission in different regions of the galaxy, we find a clear excess of [CII] emission along the impact front with the intra-group medium. We estimate that the shock due to the impact with the intra-group medium increases the [CII] emission along the shock front by 60\% and the global [CII] emission by approximately 25\% with respect to the predicted [CII] emission assuming only excitation caused by stellar radiation. This result shows the danger of interpreting [CII] emission as directly related to star formation since shocks and other mechanisms can significantly contribute to the total [CII] emission from galaxies in groups and clusters.
\end{abstract}



\section{Introduction} \label{sec:intro}
Although galaxy clusters have historically been believed to be closed and dynamically relaxed systems at the present epoch, a large fraction of them instead are continuing to grow through the merger of subclusters and the infall of galaxies~\citep{McGee2009}, usually acquired through surrounding filaments~\citep[see, e.g., ][]{Fadda2008}. As infalling galaxies enter the diffuse hot gas which permeates clusters and massive groups~\citep[see, e.g., ][]{Sarazin1986}, they experience ram-pressure which can unbind their gas from their gravitational potential~\citep{Gunn1972}. 
This effect can eventually strip most of the gas from the galaxies, leading to the quenching of star formation~\citep{vanGorkom2004}. The affected galaxies appear to be morphologically disturbed and with trails of stripped gas~\citep{Gavazzi1995, vanGorkom2004}. In some extreme cases `Jellyfish galaxies' are observed, whose name is evocative of the tentacles of gas trailing the galaxy~\citep{Ebeling2014, Boselli2016, Sun2006}. Before the complete removal of gas, moderate values of ram pressure can lead to an increase of the star formation rate in the regions close to the impact with the intra-cluster medium~\citep{Merluzzi2013, Vulcani2018}. In fact, the increased pressure helps compress the gas and triggers more star formation~\citep{Kapferer2009}. Over time, however, the interstellar medium is fully stripped from the galaxy and star formation ceases~\citep{Bekki2009}. 

Arp~25 is a beautiful example of an infalling galaxy in the initial phase of the interaction with the intra-group medium. It resides inside a group of galaxies, the NGC~2300 group, which was the first group where X-ray emitting intra-group medium was observed~\citep{Mulchaey1993}. ROSAT observations revealed a surprisingly dense ($\approx 5.3 \times 10^{-4}$~cm$^{-3}$) and extended ($\approx0.2$~Mpc) intra-group gas halo. Under several standard assumptions, the total mass inside this region is about $3\times10^{13}$~M$_{\odot}$. The barionic mass being less than 15\%, the presence of such a large amount of gas can be explained only by invoking a large quantity of dark matter in this group. The intra-group gas is hot ($\approx 0.9$~keV) and relatively metal poor ($\approx 0.06$~Z$_{\odot}$), revealing very little loss of processed gas from member galaxies \citep{Mulchaey1993, Davis1996}.

Deeper observations with XMM \citep{Finoguenov2006} and Chandra \citep{Rasmussen2006, Wolter2015} lead to the discovery of many X-ray ultra-luminous sources in Arp~25. The data were found consistent with intra-group gas being pressurized at the leading edge due to the supersonic motion of the galaxy through the intra-group medium.  Although the ram pressure significantly affects the morphology of the outer gas disc, it is probably insufficient to strip large amounts of cold gas from the disc. According to the analysis of \citet{Rasmussen2006}, the X-ray data are consistent with a mildly shocked intra-group medium.

As this galaxy is viewed nearly face-on, the deformation of its spiral morphology caused by ram pressure is perfectly observable. 
H$\alpha$ observations of Arp~25 \citep{Tomicic2018} show a front of enhanced star formation on the leading edge and a gradient of star formation in the direction perpendicular to the impact. However, the authors were not able to identify shocks using optical line diagnostics.

Low velocity shocks have been invoked to explain the high [CII] emission detected in studies of compact groups \citep{Appleton2017, Alatalo2014} and clusters \citep{Minchin2022}. In fact,  an effective way to dissipate the energy of the shock which accumulates in the molecular hydrogen is via emission of the fine-structure [CII] line at 157.7~$\mu$m \citep{Lesaffre2013}. This line typically acts as a coolant of the warm molecular and atomic hydrogen excited by the radiation from bright young stars. Since the far-infrared (FIR) continuum is produced by the emission of dust excited by the same stars, an excess of the [CII]/FIR ratio can be used to detect shocks in the molecular hydrogen. In this paper, we show how the ram-pressure is not only triggering star formation along the impacted region but it is also responsible for shocking the interstellar medium of the galaxy in the same regions. This conclusion is based on recent photometric and spectroscopic observations in the far-IR obtained with SOFIA, the Stratospheric Observatory For Infrared Astronomy. These observations, which were performed during the last months of activity of the observatory, show the enormous potential of far-infrared studies to unveil environmental effects on the evolution of galaxies in groups and clusters.

Throughout this paper, we use $H_0$~=~70\, km\, s$^{-1}$ \, Mpc$^{-1}$, $\Omega_{\rm m}$ = 0.3, and $\Omega_{\rm \Lambda}$ = 0.7. The adopted distance of Arp~25 is discussed in Section~\ref{sec:group} and reported in Table~\ref{tab:properties}.

\begin{deluxetable}{lcc}
\label{tab:properties}
\tablecaption{General properties}
\tabcolsep=0.03cm
\tablehead{\colhead{Quantity} &\colhead{Value} &\colhead{Reference}}
\startdata
R.A.(J2000) & 07$^h$ 27$^m$ 14.36$^s$ & \\
Dec (J2000) &  +85$^o$ 45' 16.4" & \\
Luminosity Distance & 28.5 Mpc & this paper\\
Angular Distance & 28.2 Mpc & this paper\\
Scale  & 138 pc/arcsec & \\
Inclination & 20\textdegree $ \pm 10$\textdegree & \citet{Tomicic2018}  \\
v$_{sys}$ & 2416$\pm$2 km/s& \citet{Reid2019}\\
Type & SAB(rs)c  & \citet{RC3}\\
\enddata
\end{deluxetable}

\section{Data and Observations} \label{sec:data}
\label{sec:observations}

\begin{figure}[!b]
\centering\includegraphics[width=0.49\textwidth]{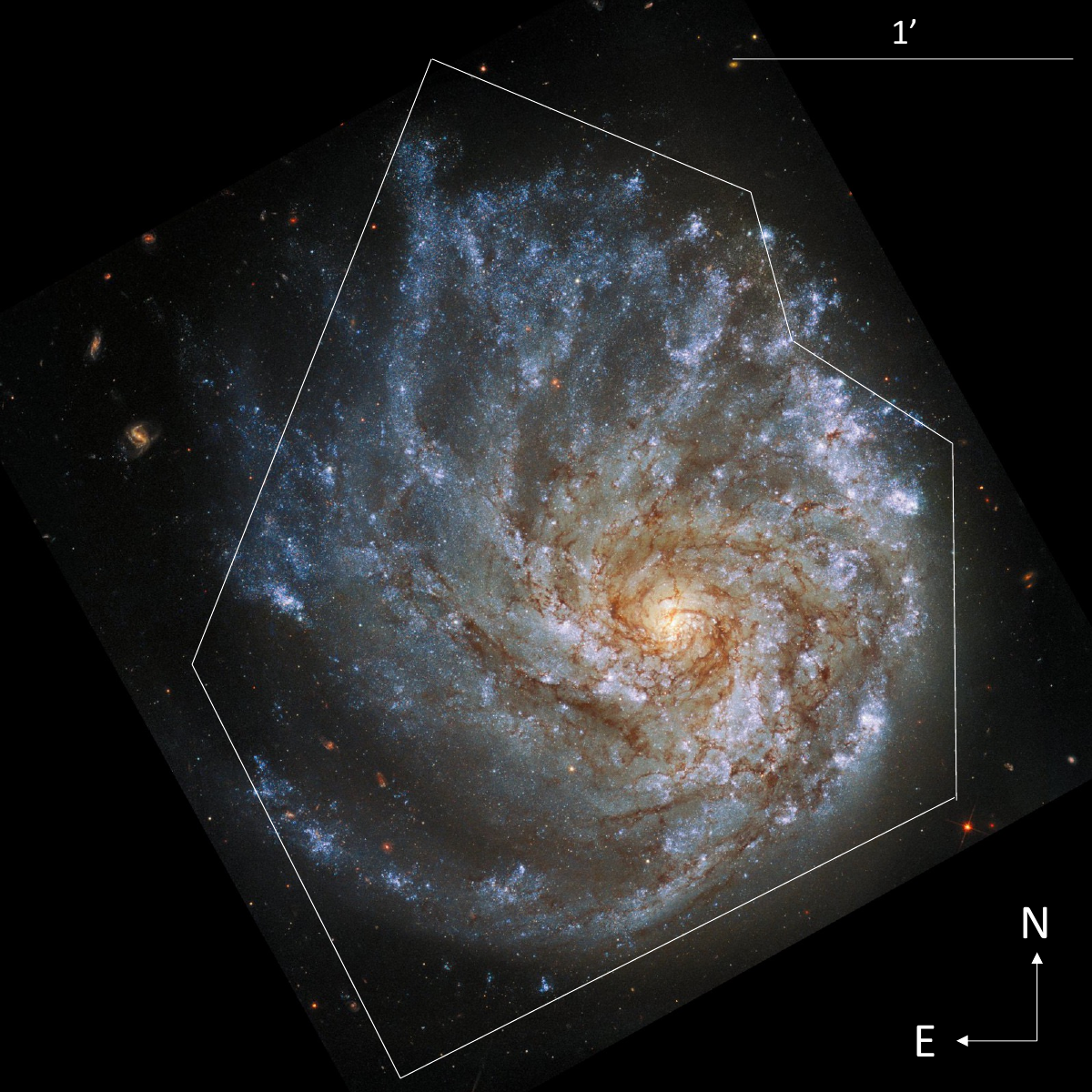}
\caption{Coverage of the FIFI-LS observations (white contour) over a composite HST image of Arp~25 obtained from WFC3 images in the 275~nm, 336~nm, 438~nm, 555~nm, and 814~nm bands. Credit: ESA/Hubble \& NASA, P. Sell, Acknowledgement: L. Shatz}
\label{fig:coverage}
\end{figure}

\begin{figure*}[!t]
\includegraphics[width=\textwidth]{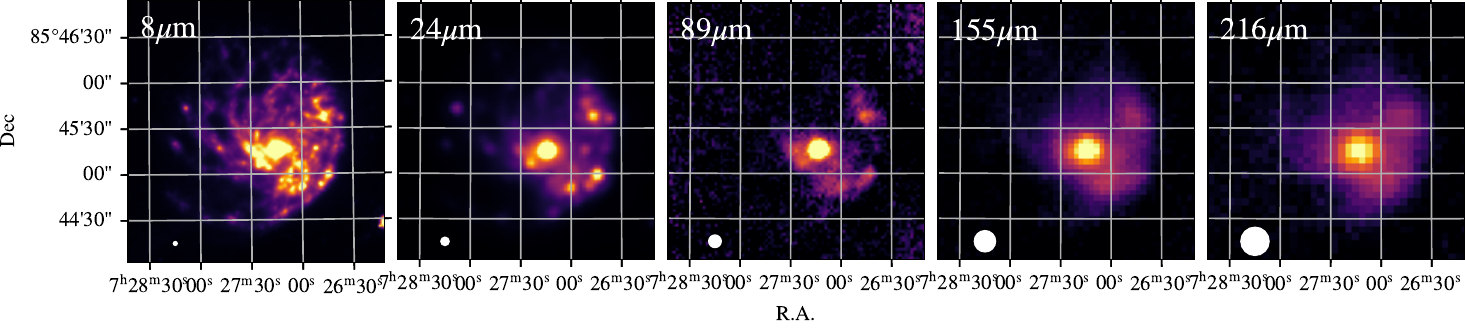}
\caption{Infrared emission from Arp~25 observed with Spitzer (8 and 24~$\mu$m) and SOFIA/HAWC+ in the C, D, and E bands. The white circles in the left-low corner of each map correspond to the beams of each observation.
}.
\label{fig:ir}
\end{figure*}

\subsection{SOFIA data}
\label{sec:SOFIA}
Data for Arp~25 were obtained during cycle~9 as part of two DDT proposals.
The first proposal used HAWC+ and was submitted as a flash proposal (P.I. Minchin) and executed in two flights (SOFIA flights 883 and 885) in June 2022. Following the good detection of the galaxy in the three HAWC+ bands (C, D, and E) a second DDT proposal was submitted to observe the galaxy during the last observational opportunity with FIFI-LS before the decommissioning of SOFIA. The proposal (P.I. Polles) was executed on flight 907 on August 30, 2022. The FIFI-LS data were obtained at a barometric altitude of 43,000~ft with a low value of zenithal precipitated water vapor (3.6~$\mu$m). The 400~s integration contours of the FIFI-LS observations are shown on the top of an HST image of Arp~25 in Fig.~\ref{fig:coverage}.

The HAWC+ data from flight 883 (bands C and D) were obtained with the detector at a temperature slightly higher than the standard value. The observations were repeated during flight 885 when the detector was operating in standard conditions. The current paper makes use of the combination of the two observations rescaled to the well calibrated flux values observed during flight 885.  The new HAWC+ data are displayed in Figure~\ref{fig:ir}.

\subsection{Archival data}
To compute the spectral energy distribution (SED) across the galaxy, we made use of several data from the ultraviolet to the mid-infrared. 
The near- and far-ultraviolet maps were obtained by the Deep Imaging Survey with GALEX  and were retrieved from the MAST archive \citep[target name PS\_NGC4258\_MOS23, obs. ID 2606460620865798144,][]{https://doi.org/10.17909/y7vg-8a63}.
At visible bands, we used Pan-STARRS maps in the g, i, r, z, and y filters \citep{Flewelling2020}. For the near-IR bands J, H, and Ks we use 2MASS data retrieved from the IRSA archive \citep{https://doi.org/10.26131/irsa122}. The WISE band 3  image at 11.3\micron\ was obtained from the all--sky survey \citep{https://doi.org/10.26131/irsa153}, while the Spitzer data were obtained from the Spitzer archive \citep{https://doi.org/10.26131/irsa413}.

Several spectral cubes have been used in our analysis of the gas in Arp~25. They come from different surveys which made their data publicly available. H$\alpha$ data are from the GHASP survey \citep{Epinat2008} and they have been obtained from the Fabry-P\'erot database~\footnote{\url{https://cesam.lam.fr/fabryperot/}} of the Observatoire de Haute Provence. CO data are from the  COMING survey~\footnote{\url{https://astro3.sci.hokudai.ac.jp/~radio/coming/data/}} \citep{Sorai2019} and were obtained with the Nobeyama 45m telescope. Finally, HI data are from the WHISP survey~\footnote{\url{www.astron.nl}} \citep{vanderHulst2001} obtained at the Westerbork telescope.

\section{Results and Discussion}
\label{sec:results}

\subsection{The NGC~2300 group and the distance of Arp~25}
\label{sec:group}

\begin{figure}[t!]
\begin{center}
\includegraphics[width=0.47\textwidth]{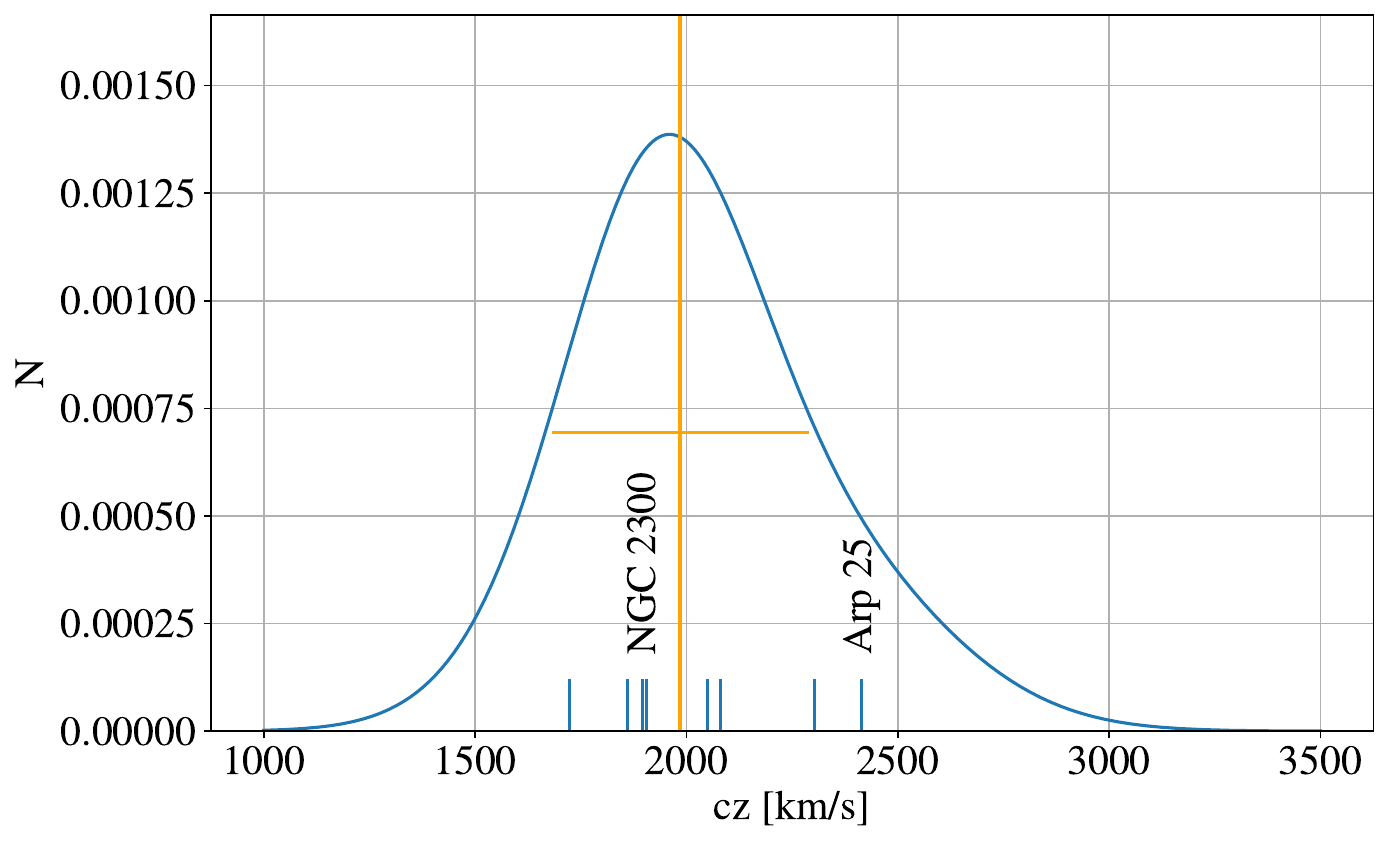}
\end{center}
\caption{
Adaptive kernel histogram of the velocity distribution of the NGC~2300 group. Velocities of single members are indicated with vertical segments. The vertical orange line and the horizontal orange segment correspond to the velocity and dispersion of the group computed with the biweight estimator (1985 and 255~km/s).
}
\label{fig:members}
\end{figure}

We gathered the velocities of various candidate members of the NGC~2300 group from literature. Although there is no dedicated spectroscopic study of this group, \citet{Diaz2012} report velocities of four galaxies in this group, while \citet{Wolter2015} considers five of them. A search for possible group members in literature with distances of less than 0.4~Mpc from the group center identified by the peak of the X-ray emission~\citep{Mulchaey1993}, and velocities between 1000 and 3000~km/s yields a total 8 members (see Table~\ref{tab:group}). Arp~25 has the most extreme velocity among these members.  It is also a late-type spiral galaxy which is generally considered to be infalling and moving on a radial orbit \citep[see, e.g.,][]{Biviano2004}.
The systemic velocity of the NGC~2300 group has been computed with a biweight mean \citep{Beers1990} yielding a value of 1985~km/s. This corresponds to a luminosity distance of 28.5~Mpc.  The histogram of the velocity distribution is shown in Fig.~\ref{fig:members} and it is obtained using an adaptive kernel estimator \citep{Fadda1998}. Location and dispersion of the adaptive kernel distribution approximately correspond to the values computed using the biweight estimator ($v=1985$~km/s and $\sigma_v = 255$~km/s). The difference is probably due to the skewness of the distribution caused by the high proper velocity of Arp~25.

By accepting 1985~km/s as group systemic velocity, Arp~25 would have a line-of-sight velocity relative to the group of 430~km/s. In the worst-case scenario, the impact would be perpendicular to the plane of the disk as suggested by the symmetric deformation of the spiral morphology. Since the inclination of the galaxy inferred from the rotational velocity is approximately 20\textdegree\ \citep{Tomicic2018}, this would corresponds to an infall velocity of approximately $v_{infall} = 430/\sin(20^{\circ}) \approx 1260$~km/s. This velocity is sufficient to generate enough ram pressure in a low density medium similar to the intra-group medium in the NGC~2300 group to explain the shock seen in the X-ray observations. \citet{Rasmussen2006}, on the basis of the observed shock, estimate a velocity of $860\pm120$~km/s but they do not exclude a higher value because of the lack of knowledge about the three-dimensional direction of the motion of the galaxy.

\begin{deluxetable}{lcccc}[!t]
\label{tab:group}
\tablecaption{NGC 2300 group members}
\tabcolsep=0.02cm
\tablehead{
\colhead{Name} &
\colhead{R.A. - Dec (J2000)} &
\colhead{v [km/s]} &
\colhead{R [Mpc]} &
\colhead{Src}}
\startdata
X-ray center& 07:30:39.54  +85:40:59.0 &     -        & -   & 0\\
NGC 2300    & 07:32:20.49  +85:42:31.9 & 1905 $\pm$  7&  2.4' [0.02]& 1,5\\
Arp 25      & 07:27:14.36  +85:45:16.4 & 2416 $\pm$  2&  5.7' [0.05]& 1\\
IC 455      & 07:34:57.53  +85:32:13.9 & 2050 $\pm$ 51& 10.0' [0.08]& 1,5\\
UGC 3670    & 07:20:04.73  +85:35:14.3 & 1861 $\pm$ 29& 13.4' [0.11]& 3\\
UGC 3654    & 07:17:47.09  +85:42:47.7 & 2303 $\pm$ 22& 14.6' [0.12]& 1\\
CGCG 362-035& 07:15:06.84  +85:46:28.4 & 1724 $\pm$ 30& 18.2' [0.15]& 1\\
CGCG 362-048& 07:58:12.74  +85:43:00.0 & 1896 $\pm$ 29& 31.0' [0.26]& 3\\
IC 469      & 07:55:59.08  +85:09:32.1 & 2080 $\pm$ 39& 43.6' [0.36]& 4
\enddata
\tablecomments{Sources code for the last column: 0 \citep{Mulchaey1993}, 1 \citep{Huchra2012}, 2 \citep{Springob2005}, 3 \citep{Falco1999}, 4 \citep{RC3}, 5 \citep{Afanasiev2016}.}
\end{deluxetable}

By means of the virial theorem it is possible to estimate the mass of the group from positions and velocities of the galaxy members. We adopt the systemic velocity of $v_S = 1985$~km/s and compute the three-dimensional velocity dispersion with the formula:

\begin{equation}
    v^2 = \frac{3}{n} \sum_i \frac{(v_i - v_S)^2 - v_{err}^2}{(1-\frac{v_i v_S}{c^2})^2},
\end{equation}

where $v_{err}^2 = \sum \sigma_v^2$ is the quadratic sum of the errors on the galaxy velocities, $n$ is the number of galaxies, and $c$ is the speed of light. The denominator takes into account the relativistic correction, the factor 3 allows one to pass from line-of-sight velocities to the three-dimensional distribution, and the subtraction of the error term compensate for the broadening of the velocity distribution due to measurement errors \citep{Danese1980}.

The projected virial radius $R_{vir}$ is estimated using the formula (3) from \citet{Carlberg1996} assuming the diffuse X-ray center of emission as the center of the cluster (see Table~\ref{tab:group}) and equal weights for all the galaxies:
\begin{eqnarray}
R_{vir}^{-1} &=& \frac{1}{n^2} \sum_{i<j} \frac{1}{2\pi} \int_0^\pi [r_i^2 + r_j^2 + 2 r_i r_j \cos{\theta}]^{-1/2} d\theta \\ \nonumber
&=& \frac{2}{\pi n^2} \sum_{i<j} \frac{1}{(r_i+r_j)} \int_0^{\pi/2} [1 - m_{ij} \sin^2 t]^{-1/2} dt\\\nonumber
&=&\frac{2}{\pi n^2} \sum_{i<j} \frac{K(m_{ij})}{(r_i+r_j)}.
\end{eqnarray}
The formula can be expressed as a complete elliptical integral of the first type $K(m_{ij})$ with $m_{ij}=\frac{4 r_i r_j}{(r_i+r_j)^2}$ which can be computed with the ellipk function in the Python scipy library \citep{2020SciPy}. The three-dimensional virial radius, $r_{vir}$, is obtained from $R_{vir}$ with a deprojection factor $r_{vir} = \pi R_{vir}/2$ \citep{Limber1960}.\\
The computation of the virial mass:
\begin{equation}
    M_{vir} = \frac{r_{vir} v^2}{G},
\end{equation}
with $G$, the gravitational constant, yields a value of $(1.8 \pm 0.3) \times 10^{13}$~M$_{\odot}$, obtained by resampling the data with the bootstrap technique \citep[see, e. g., ][]{Efron1982}. This value is close to that obtained from the X-ray diffuse emission by \citet{Mulchaey1993} and confirms the large amount of dark matter needed to explain the stability of this group.

\begin{figure*}[!t]
\centering\includegraphics[width=0.95\textwidth]{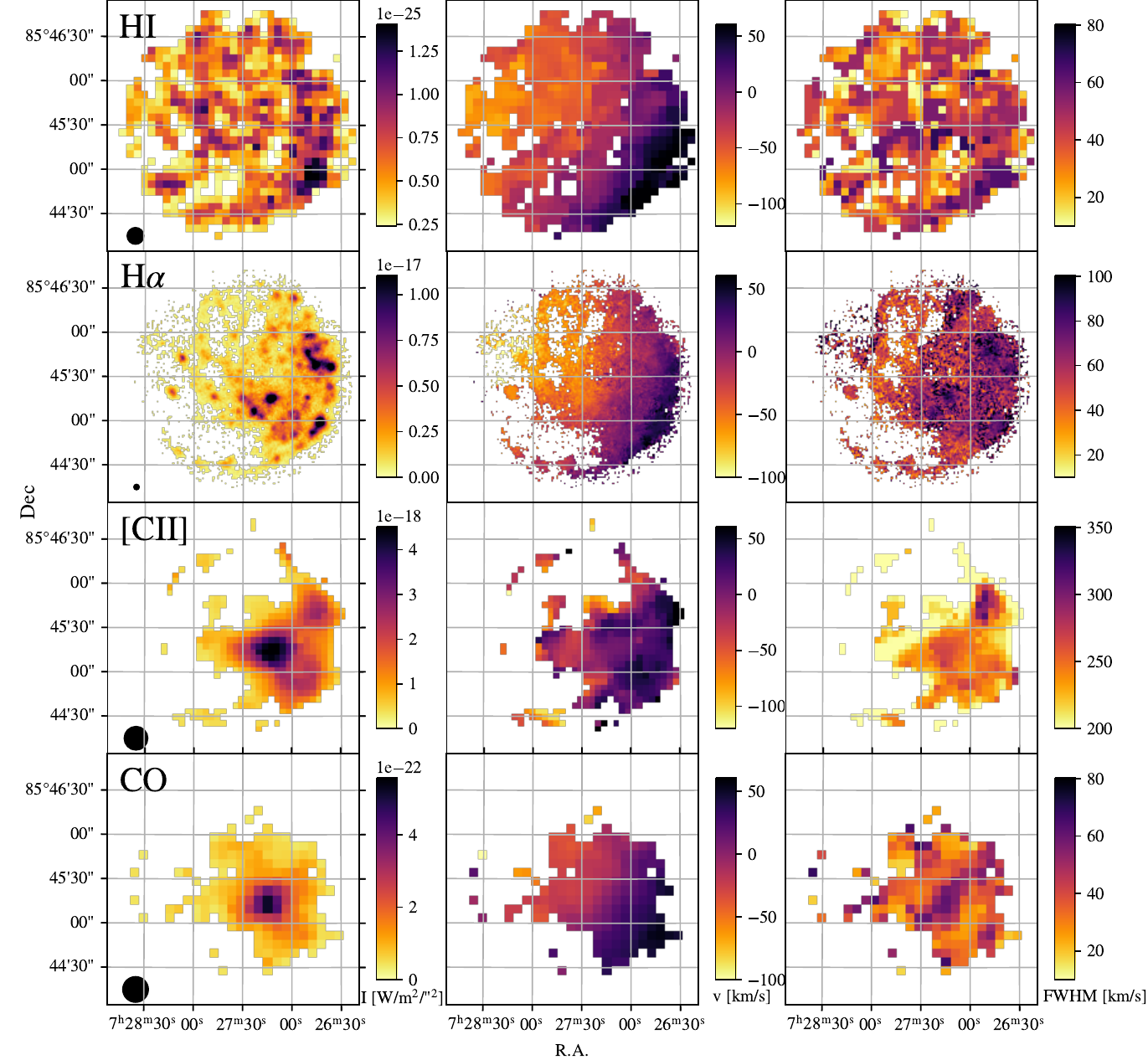}
\caption{Intensity, velocity, and velocity dispersion maps for the HI, H$\alpha$, [CII], and CO lines of Arp~25. These lines show the distribution of neutral and ionized atomic hydrogen (HI and H$\alpha$) and of the warm and cold molecular hydrogen ([CII] and CO). The beam of each observation is shown as a black circle in the bottom left corner of each intensity map. The images show the lopsided emission from the ionized atomic and warm molecular hydrogen, while the neutral atomic and cold molecular gas emission is more uniform across the galaxy. A gradient in velocity is clearly visible, while a somewhat enhanced velocity dispersion is visible along the shock region.
}
\label{fig:moments}
\end{figure*}

\subsection{Dust emission maps}
\label{sec:dust}

Figure~\ref{fig:ir} shows the emission of the dust in the mid- and far-IR as seen by Spitzer at 8 and 24~$\mu$m with IRAC and MIPS, respectively, and by SOFIA/HAWC+ in the C, D, and E bands which correspond to central wavelengths of 89, 155, and 216~$\mu$m. In all the maps the nucleus is the brightest peak of emission. Three other peaks are visible in all the maps along the shock front, although the two peaks in the southern part almost merge in the images at longer wavelengths (band D and E). These images confirm the excess of star formation on the side affected by ram pressure seen by \citet{Tomicic2018} with H$\alpha$ imaging (see also Section~\ref{sec:SF}).

\subsection{Moment maps}
\label{sec:maps}

In Figure~\ref{fig:moments} we compare the intensity, velocity, and velocity dispersion maps of the HI, CO, [CII], and H$\alpha$ observations of Arp~25. These observations map the main states of the atomic and molecular gas in Arp~25: the neutral atomic gas (HI), the cold molecular gas (CO), the warm molecular gas ([CII]) and the ionized atomic gas (H$\alpha$). 
It is immediately evident that the intensity maps of H$\alpha$ and [CII] are lopsided. The shock front is well recognizable. The intensity map of the CO and HI emission are more uniform, although the HI has an excess of emission along the shock front. In the CO map, the peak of the intensity corresponds to the nucleus and there are no comparable peaks of emission along the shock front. The nucleus has very low emission in the HI map as is usually the case in star-forming galaxies. The velocity gradient is well visible in all the maps. Finally, the velocity dispersion maps show higher values along the shock front and, except for the HI map, on the nucleus of the galaxy.

The difference of emission along the shock front between the [CII] and the CO observations is remarkable. Clearly the source of emission is different in the two cases. The two peaks of [CII] emission along the shock front corresponds to the most intense spots in the H$\alpha$ map. However, as we will see in the following, the ratio between the [CII] emission and the dust emission is much higher than that expected for normal star formation.

\begin{deluxetable}{lcccccc}
\label{tab:sedbands}
\renewcommand{\arraystretch}{1.0}
\tabcolsep=0.06cm
\tablecaption{Bands used for SED Fitting}
\tablehead{\colhead{Filter} &\colhead{Wavelength}&\colhead{Beam}&\colhead{Pixel}&\colhead{$\sigma_{\rm{cal}}$} &\colhead{Ext}& \colhead{Refs} \\ 
\colhead{Name} & \colhead{ $\mu$m } &\colhead{arcsec}&\colhead{arcsec}& \colhead{mag} &\colhead{mag} & \colhead{}} 
\startdata
GALEX\_FUV & 0.1516 &4.2&1.5& 0.05  & 0.70& 1\\
GALEX\_NUV & 0.2267 &5.3&1.5& 0.03  & 0.79& 1 \\
PANSTARRS g& 0.4866&1.3&0.258&0.020 & 0.30&2\\
PANSTARRS r& 0.6215&1.2&0.258&0.016 & 0.23&2\\
PANSTARRS i& 0.7475&1.1&0.258&0.017 & 0.18&2\\
PANSTARRS z& 0.8679&1.1&0.258&0.018 & 0.13&2\\
PANSTARRS y& 0.9633&1.0&0.258&0.022 & 0.11&2\\
2MASS\_$J$ & 1.235 &2.9&2.0& 0.03   & 0.08&3 \\
2MASS\_$H$ & 1.662 &2.8&2.0& 0.03   & 0.05&3 \\
2MASS\_$Ks$& 2.159 &2.9&2.0& 0.03   & 0.03&3 \\
IRAC\_1    & 3.550 &1.66&1.2& 1.8\% & &4 \\
IRAC\_2    & 4.490 &1.72&1.2& 1.9\% & &4 \\
IRAC\_3    & 5.730 &1.88&1.2& 2.0\% & &4 \\
IRAC\_4    & 7.870 &1.98&1.2& 2.1\% & &4 \\
WISE\_3    & 12.08 &6.5&2.75& 4.5\% & &5\\
MIPS\_24   & 23.70 &4.9&2.5& 4.0\%  & &6\\
HAWC+ C    & 89    &7.8&4.0&10\%    & &7 \\
HAWC+ D    & 154  &13.6&6.9&10\%    & &7 \\
HAWC+ E    & 214  &18.2&9.4&10\%    & &7 \\
\enddata
\tablecomments{Beam sizes for SDSS and 2MASS are median seeing values. References: (1) \citet{Morrissey2007}, (2) \citet{Tonry2012, Magnier2020}, (3) \citet{Skrutskie2006}, (4) \citet{Reach2005}, (5) \citet{Jarrett2011}, (6) \citet{Engelbracht2007}, (7) \citet{Harper2018}. Extinction mags are computed according to \citet{Cardelli1989} and the $A_v = 0.2663$ value from \citet{Schlafly2011}.}
\end{deluxetable}

\subsection{Apertures}
\label{sec:regions}

To better study the effect of the ram pressure on the [CII] emission, we defined a series of independent apertures centered on the regions with [CII] and/or H$\alpha$ emission. In particular, we defined an aperture centered on the nucleus of the galaxy, five apertures along the shock front, three apertures in the region between the nucleus and the shock front (which we will call post-shock region) and other seven apertures in other regions of the galaxy with far-IR emission.
The apertures have a diameter of 18~arcsec which corresponds to the beam of the HAWC+ band E, the band considered for obtaining the spectral energy distribution with the poorest spatial resolution. 
The apertures are reported in the top panel of Fig.~\ref{fig:apertures} with four different colors: lime green for the shock region, green for the post-shock region, blue for the nucleus, and orange for the disk regions. The same color code is used in the plots of the following sections.
Before performing aperture photometry, we took care of removing any residual background from the optical and near-IR images. The 2MASS images had a gradient in the background.
To remove it, we first masked the region with extended emission around Arp~25 (a disk of 3~arcmin diameter) and all the point sources in the field. Then, we computed the median flux for each column of the image and smoothed the obtained background profile with a Chebyshev polynomial. This passage allowed us to avoid adding noise when removing the residual background from the image. The archival PanSTARRS stacked images contain several artefacts and have an uneven background. To improve the quality of the stacked images, we downloaded all the single images used to obtain the archival stacks (called ``warp" images). After discarding the images with bad seeing or with too many artefacts,  we masked a few remaining artefacts, subtracted the residual background from each single exposure, and stacked the selected images scaled to the same photometric zero-point. In this way we obtained cleaner images with a more even background.

\subsection{SED Modeling}
\label{sec:SED}

\begin{figure}
\begin{center}
\hspace*{-0.4cm}\includegraphics[width=0.49\textwidth]{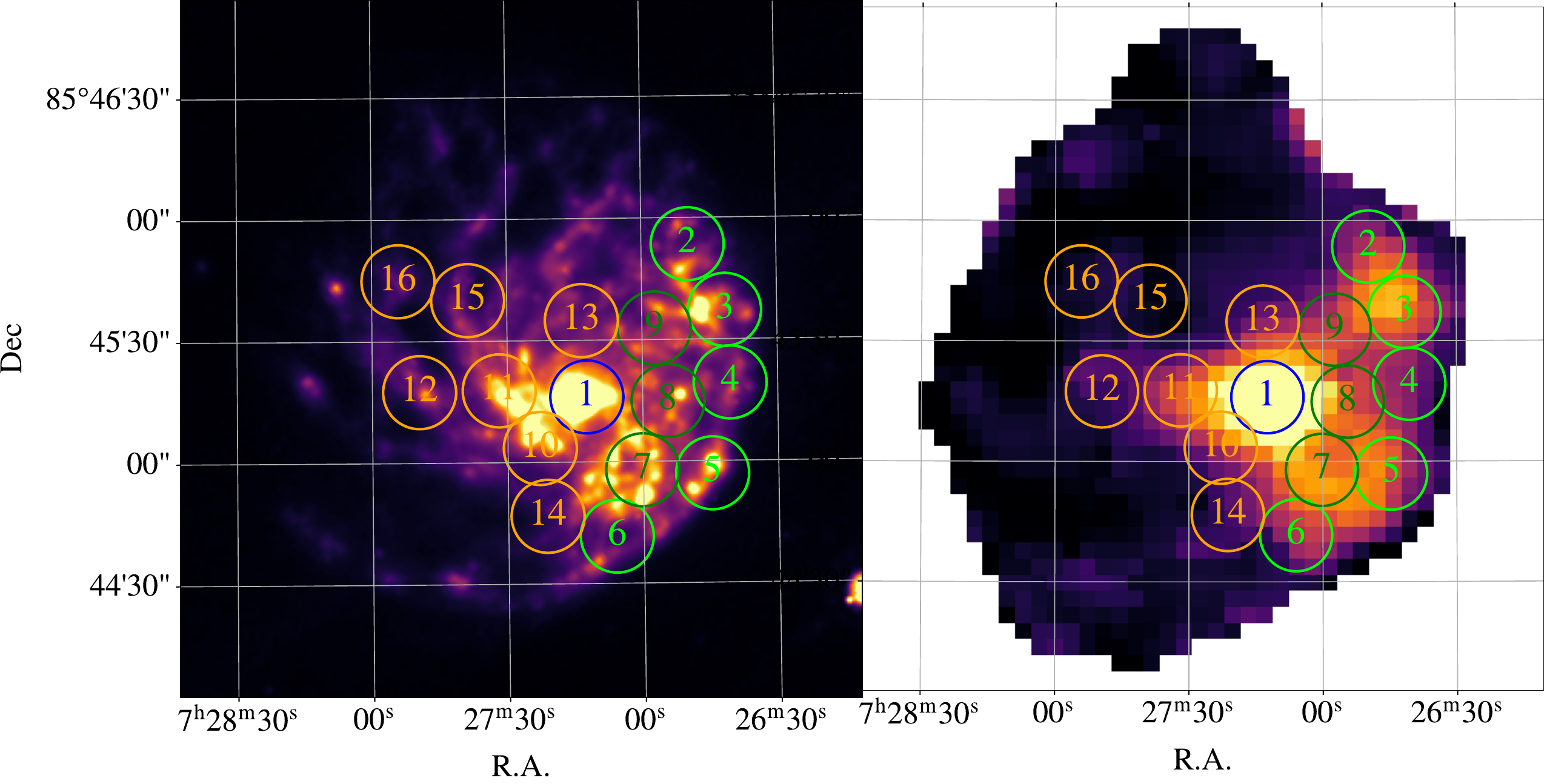}\\
\includegraphics[width=0.49\textwidth]{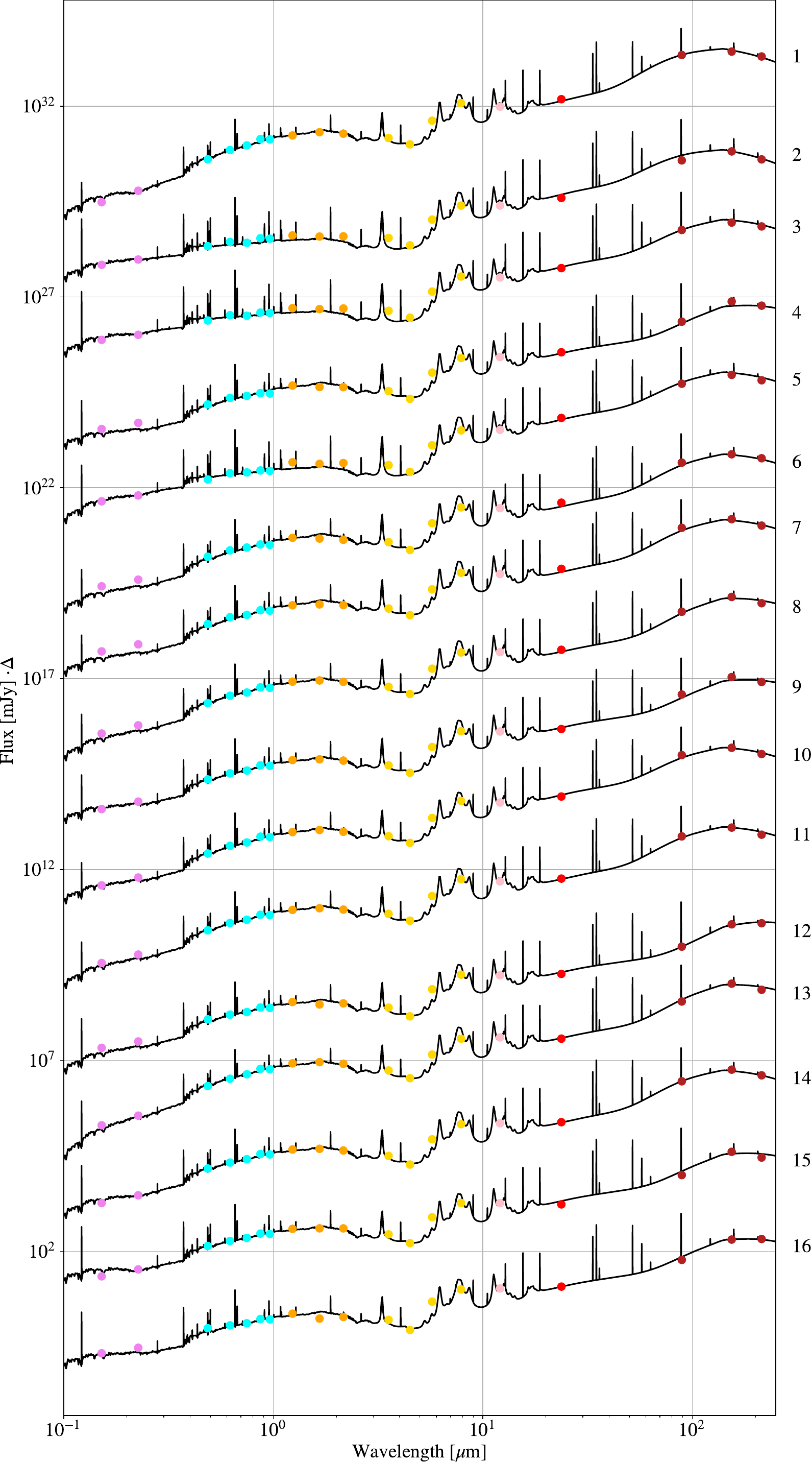}
\end{center}
\caption{{\it Top}: Circular apertures selected for the study of the different parts of the galaxy overlapped on the 8$\mu$m (left) and [CII] images (right). Colors are as in Fig.~\ref{fig:KS}. {\it Bottom}: SED for the apertures, multiplied by a factor $\Delta = 100^{16-i}$ where $i$ is the identification number of the aperture shown on the right side of the plot.
}
\label{fig:apertures}
\end{figure}

Once all photometric data had been smoothed to the same spatial resolution as the HAWC+ E band and the aperture fluxes measured, Spectral Energy Distribution (SED) fitting was performed using the Code Investigating GALaxy Evolution \citep[CIGALE][]{Noll2009, Boquien2019}. CIGALE was chosen over other SED modeling tools \citep[i.e., MAGPHYS, ][]{daCunha2008} because of the ease of adding filter profiles, namely the HAWC+ bands which are not supported in the latest version of MAGPHYS.  It should be noted that there has been extensive work comparing CIGALE and other SED modeling tools, yielding no significant differences \citep{Hunt2019}. CIGALE models the SED by assuming the energy absorbed by dust from the UV to the near--infrared is balanced by the energy emitted by dust in the mid and far--infrared.  The full suite of photometric observations used to determine the SED fits are described in Section~\ref{sec:observations} and are listed in Table~\ref{tab:sedbands} along with the calibration uncertainties used to estimate the errors.   Fits were determined using the \citet{BruzualCharlot2003} stellar population and the \citet{Draine2014} dust models.  A complete list of the parameters and modules used in the SED models can be found in Table~\ref{tab:sedprops}.  SED fits were performed on the apertures discussed in Section~\ref{sec:regions}. Errors were determined using the sum in quadrature of the variation of the sky brightness and the calibration uncertainty for each photometric detector listed in Table~\ref{tab:sedbands}. The best model SEDs are plotted in the bottom panel of Figure~\ref{fig:apertures} as black lines overlapped on the photometric data. The color of the points refer to different instruments: GALEX in purple, PanStarrs in cyan, 2MASS in orange, IRAC in yellow, WISE in pink, MIPS in red, and HAWC+ in brown.
These SED models allow for estimations of the dust properties and attenuation rates in different environments in Arp~25.

\begin{deluxetable}{ll}
\tablecaption{Parameter values for CIGALE modules}
\label{tab:sedprops}
\renewcommand{\arraystretch}{0.87}
\tablehead{\colhead{Parameter} &\colhead{Input values} } 
\startdata
\multicolumn{2}{c}{\texttt{sfhdelayed}}\\
\texttt{tau\_main} [Gyr] & [0.5, 10], $\delta=0.25 $ \\
\texttt{age\_main} [Gyr] & 11 \\
\texttt{tau\_burst} [Gyr] & 0.05 \\
\texttt{age\_burst} [Gyr] & 0.02 \\
\texttt{f\_burst} & 0 \\
\texttt{SFR\_A} [M$_{\odot}$/yr] & 1 \\
\hline
\multicolumn{2}{c}{\texttt{bc03}}\\
\texttt{imf} & 1 (Chabrier) \\
\texttt{metallicity} [solar] & 0.02 \\
\texttt{seperation\_age} [Gyr] & 0.01 \\
\hline
\multicolumn{2}{c}{\texttt{nebular}}\\
\texttt{logU} & -3 \\
\texttt{f\_esc} & 0 \\
\texttt{f\_dust} & 0 \\
\texttt{lines\_width} [km s$^{-1}$] & 300 \\
\hline
\multicolumn{2}{c}{\texttt{dustatt\_modified\_starburst}}\\
\texttt{E\_BV\_nebular} [mag] & [0,1.0], $\delta=0.1$\\
\texttt{E\_BV\_factor} & 0.44 \\
\texttt{uv\_bump\_wavelength} [nm] & 217.5 \\
\texttt{uv\_bump\_width} [nm] & 35 \\
\texttt{uv\_bump\_amplitude} & 0, 1.5, 3 (Milky Way) \\
\texttt{powerlaw\_slope} & [-0.5, 0.0], $\delta=0.1$  \\
\texttt{Ext\_law\_emission\_lines} & 1 (Milky Way) \\
\texttt{Rv} & 3.1 \\
\texttt{filters} & B\_B90, V\_B90, FUV \\
\hline
\multicolumn{2}{c}{\texttt{dl2014}}\\
\texttt{qpah} & 0.47, 2.50, 4.58, 6.63 \\
\texttt{umin} & 0.1, 0.25, 0.5, 1,\\
& 2.5, 5, 10, 25 \\
\texttt{alpha} & 2 \\
\texttt{gamma} & 0.001, 0.002, 0.004,\\
& 0.008, 0.016, 0.032, \\
& 0.064, 0.125,0.25, 0.5\\
\hline
\multicolumn{2}{c}{\texttt{fritz2006}}\\
\texttt{fracAGN} & 0.0 \\
\hline
\multicolumn{2}{c}{\texttt{restframe\_parameters}}\\
\texttt{beta\_calz94} & False \\
\texttt{D4000} & False \\
\texttt{IRX} & False \\
\texttt{EW\_lines} & 500.7/1.0 \& 656.3/1.0 \\
\texttt{luminosity\_filters} & FUV \& V\_B90 \\
\texttt{colours\_filters} & FUV-NUV \& NUV-r$\prime$ \\
\hline
\multicolumn{2}{c}{\texttt{redshifting}}\\
\texttt{redshift} & 0 \\
\enddata
\end{deluxetable}

\subsection{
Ram pressure impact on star formation
}
\label{sec:SF}

\begin{figure}
\begin{center}
\includegraphics[width=0.48\textwidth]{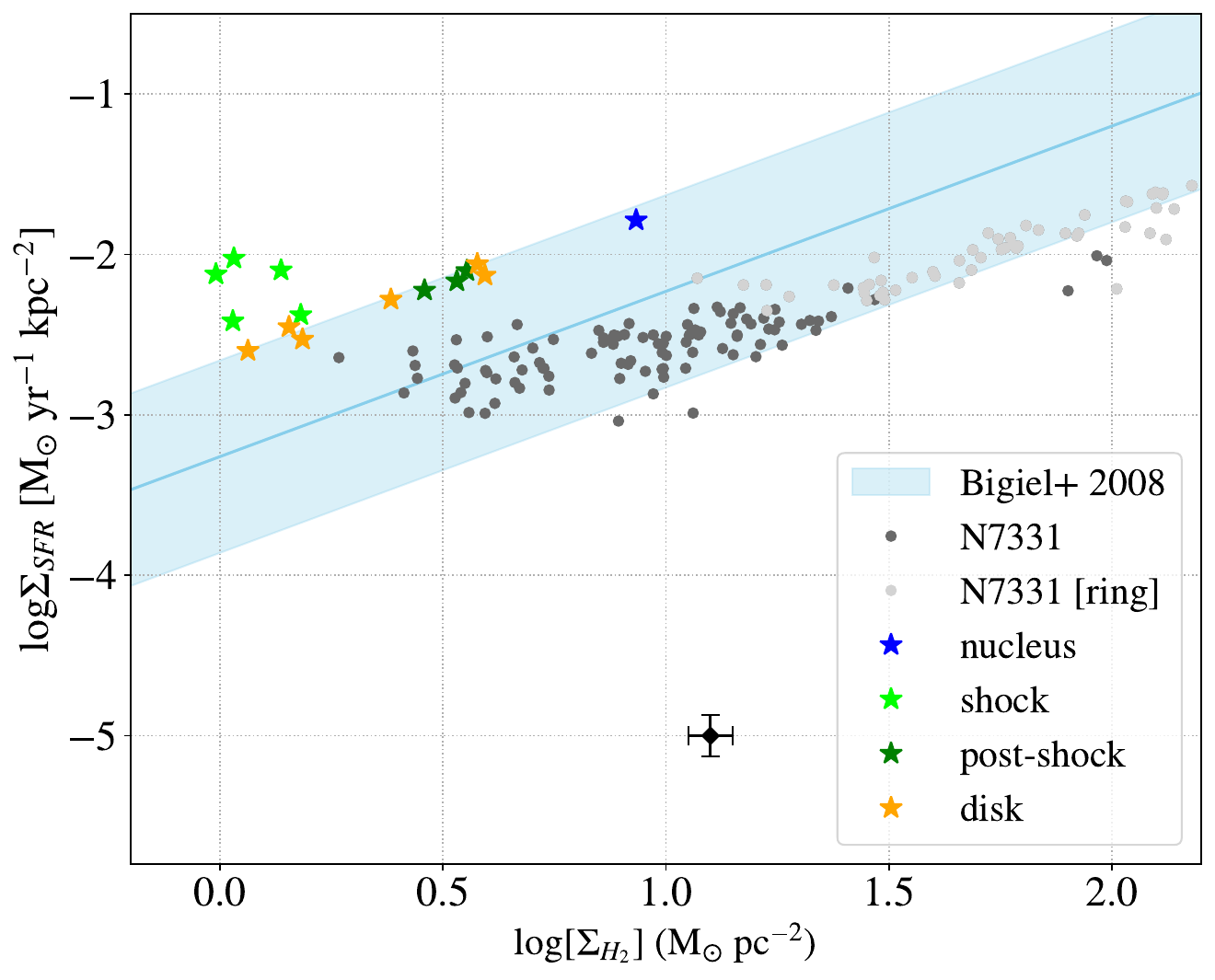}
\end{center}
\caption{Kennicutt-Schmidt diagram for the different regions in Arp~25. The values from NGC~7331, a Milky-Way analog from \citet{Sutter2022}, and the band defined in \citet{Bigiel2008} for sub-kpc regions in galaxies are plotted for comparison. The cross shows the typical error bars for the measured values of Arp~25.}
\label{fig:KS}
\end{figure}

In Figure~\ref{fig:KS} we plot the surface density of star formation ($\Sigma_{\rm{SFR}}$) versus the surface density of molecular hydrogen ($\Sigma_{H_2}$). This plot is commonly referred to as a Kennicutt--Schmidt plot, and the proportional trend between $\Sigma_{\rm{SFR}}$ and $\Sigma_{H_2}$ demonstrates how gas in the ISM is the fuel for ongoing star formation \citep{Kennicutt2012}.  

For SFR values we used the CIGALE estimates which take into account the entire SED from the far-IR to the far-UV. The H$_2$ gas mass is determined by converting the CO luminosity inside each region with Eq. 3 from \citet{Bolatto2013}.  Both the H$_2$ gas mass and the SFR are converted to surface densities by dividing by the deprojected area of the aperture used in the measurement. For comparison, we plot the relationship from \citet{Bigiel2008} which investigated the Kennicutt-Schmidt relationship at sub-Kpc scale for several star-forming galaxies as a blue line, with the dispersion represented using blue shading, as well as the relationship for regions of the Milky Way analog NGC~7331 \citep{Sutter2022}.
Figure~\ref{fig:KS} clearly shows that the star formation activity is in general high in all the regions analyzed. Along the shock front, where ram pressure is triggering further star formation, the values are substantially higher than those found in normal galaxies.
We can also notice in Figure~\ref{fig:apertures} that the stellar component in the optical side of the SED of the regions along the shock front is flatter than those of the disk regions. This confirms the formation of a younger stellar population triggered by the shock.
Although the regions along the shock front are outliers in the Kennicutt-Schmidt diagram, the relationship between star-formation rates and gas surface densities can be linearized by normalizing the gas surface density with the freefall time of the gas. In particular, \citet{Salim2015} showed that it is important to consider density dependent timescales which take into account the clumpy nature of the clouds. The relationship between a single freefall timescale at a mean density and a ``multi freefall'' timescale is a function of density variance of the clouds which can be parameterized by the sonic Mach number $\mathcal{M}$, the turbulent driving parameter $b$, and the thermal to magnetic pressure ratio $\beta$. By using the same approximations as \citet{Salim2015}, i.e. $b=0.4$ and $\beta \rightarrow \infty$, and assuming a constant ratio between $\Sigma_{gas}$ and $\Sigma_{H_2}$, we can use the logarithmic distance along the y-axis of the points in Fig.~\ref{fig:KS} from the linear relationship, $\Delta$, to roughly estimate the Mach number of the clouds in different regions of the galaxy: 
\begin{equation}
    \mathcal{M} \approx \frac{\sqrt{e^{\frac{8}{3}\Delta}-1}}{0.4}
\end{equation}
Most of the points have $\Delta = 0.5-0.6$ which corresponds to a Mach number range of 4-5, while the most shocked regions have $\Delta = 0.8 - 1.2$ corresponding to a Mach number range of 7-12. Such estimates are compatible with those reported in Table~3 of \citet{Salim2015} for local disk and starburst galaxies, respectively.

\subsection{[CII] and dust}

Most of the [CII] emission originates in the photo-dissociation region \citep[PDR, see e.g., ][]{Croxall2012} where the singly ionized carbon acts as the main coolant for the neutral gas heated by the radiation of young, bright stars. If the PDRs are in thermal equilibrium, the [CII] emission should therefore trace the star formation and be proportional to tracers of heating, such as the amount of UV attenuation, the far--infrared luminosity emitted by dust, and the intensity of the emission of polycyclic aromatic hydrocarbons (PAHs). If some region has an anomalous ratio between the [CII] emission and one of these quantities, we can deduce that some other mechanism contributes to the [CII] emission. The [CII]/UV attenuation, [CII]/FIR, and [CII]/PAH ratios have been also proposed as tracers of the photo-electric heating efficiency \citep{Kapala2017, Croxall2012}. The comparison with other normal star forming galaxies can inform us about the peculiar conditions in Arp~25.

\subsubsection{Photo-electric efficiency}

The ratio between absorbed UV radiation and emitted \CII{} defines the so-called photo-electric efficiency. We estimate the UV attenuation through the fit of SEDs with the CIGALE code. The slope of the relationship between the UV attenuation and the \CII{} surface luminosity is the photo-electric efficiency.

\begin{figure}
\begin{center}
\includegraphics[width=0.48\textwidth]{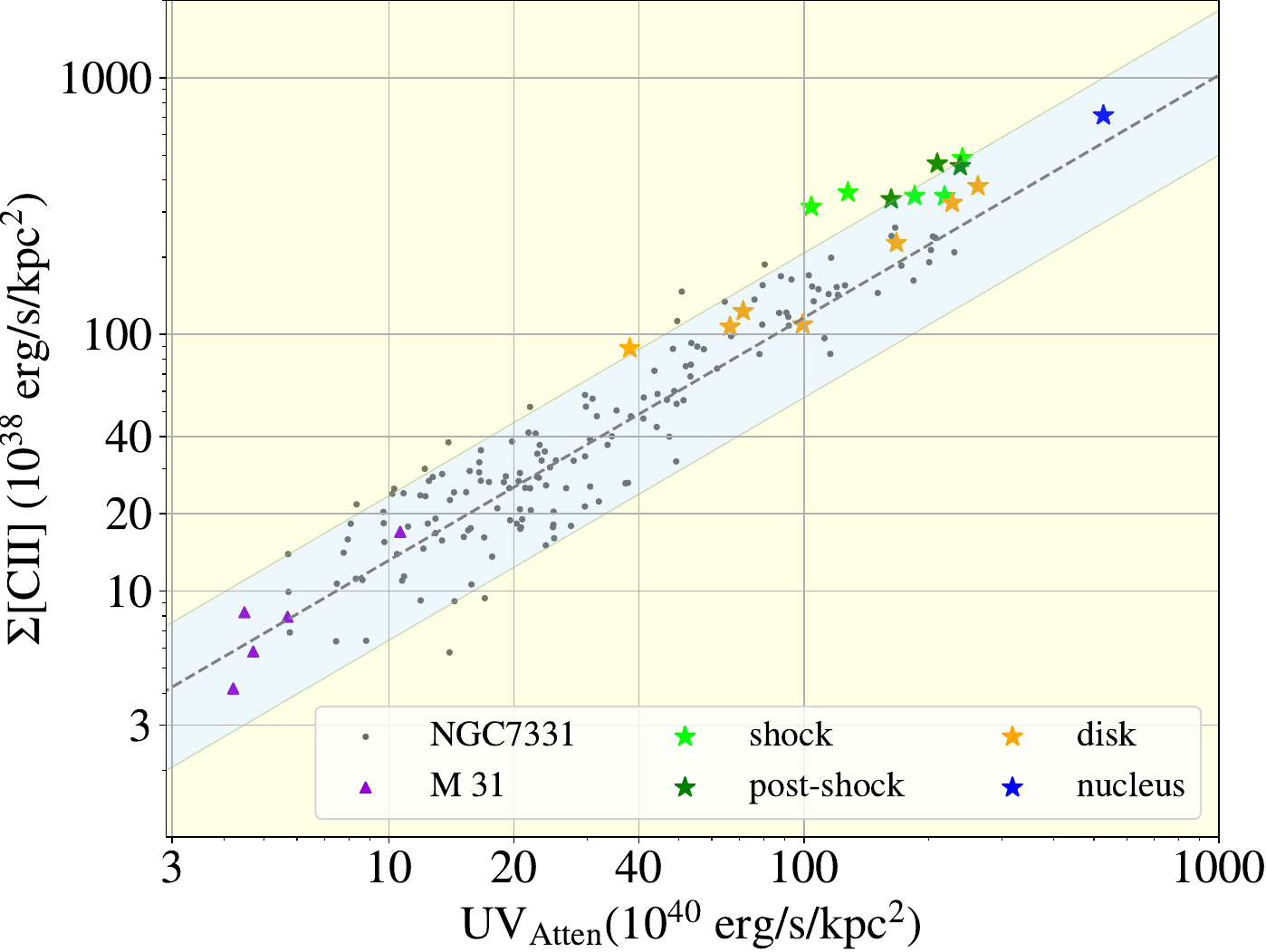}
\end{center}
\caption{Surface brightness of the [CII] line plotted as a function of the attenuated UV light determined using the CIGALE SED fits. Arp~25 data are color--coded based on their regions.  Comparison data from the nearby galaxy NGC~7331 and M~31 are shown as grey and purple symbols, respectively. A linear fit of the relationship using the NGC~7331 and M~31 and the 3-$\sigma$ region are shown shaded in light blue.}
\label{fig:UV_Atten}
\end{figure}

Figure~\ref{fig:UV_Atten} shows the relationship for two sets of reference galaxies: NGC~7331 from \citet{Sutter2022} and M~31 from \citet{Kapala2015}.
We fitted a linear relationship considering the NGC~7331 and M~31 data and assuming the line will pass through the origin. The slope of the fits is $0.96\pm0.03$\%. On each side of the relationship, we computed the dispersion of the residuals to define the region including most of the points. The 3-$\sigma$ region defined in this way is shaded in blue in Fig.~\ref{fig:UV_Atten}. While the nucleus and the disk regions follow the relationship very well and fall completely within the blue shaded region, the shock and post-shock regions show an excess of [CII] emission. In particular, the shock regions lie at 3~$\sigma$ or more above the relationship. We interpret this as evidence of the non-stellar origin of the excess of \CII{} emission in the shock regions.

\subsubsection{[CII] and dust continuum}

\begin{figure*}
\begin{center}
\includegraphics[width=\textwidth]{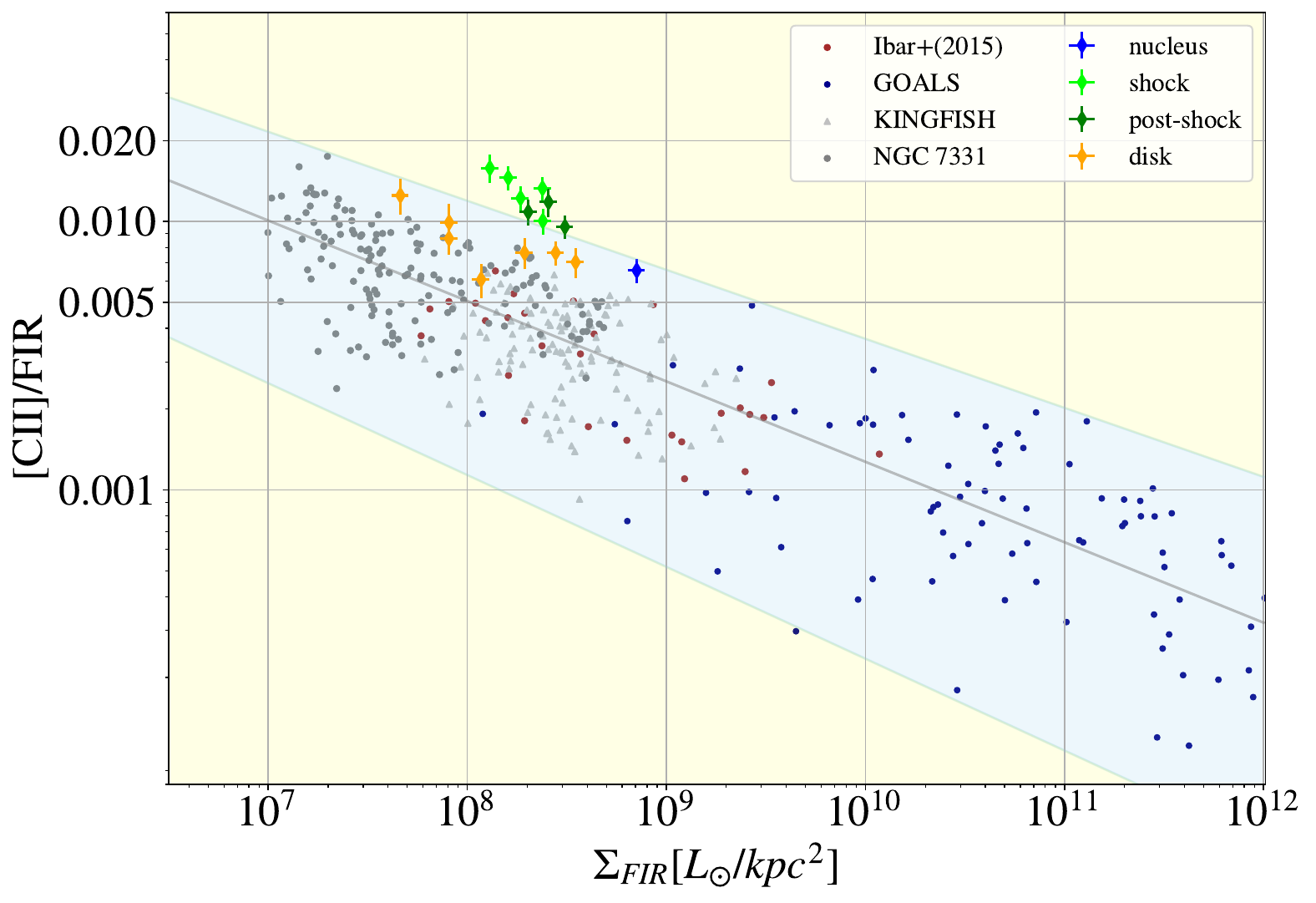}
\end{center}
\caption{The [CII]/FIR ratio plotted as a function of far-IR surface brightness ($\Sigma_{FIR}$). The points are color coded according to their region. The comparison data are from NGC~7331~\citep{Sutter2022}, star--forming regions in local galaxies from galaxies in the KINGFISH survey \citep[][]{Sutter2019}, $z\sim0.02 - 0.2$ galaxies from \citet{Ibar2015}, local U/LIRGS from the GOALS survey \citep[][]{DiazSantos2017}. The central grey line corresponds to the linear fit of the comparison points, while the lines limiting the blue shaded region correspond to 5 times the dispersion of the positive and negative residuals. Such a region contains points where the \CII{} emission is powered by star formation. Points higher than the upper line have an excess of \CII{} emission.}
\label{fig:CIIratio}
\end{figure*}

The dust emission accounts for the peak in the far-infrared of the spectral energy distribution of the galaxy. The typical estimate of this energy is the integrated FIR flux between $8$ and $1000 \mu$m, also called total infrared flux. Figure~\ref{fig:CIIratio} shows the relationship between the [CII]/FIR ratio and the FIR surface brightness.  The FIR surface brightness was determined by dividing the FIR luminosity by the deprojected area of one region.  To compare the regions within Arp~25 to previous studies of the \CII/FIR relationship, we also plot data from resolved regions across the disk of the nearby star--forming galaxy NGC~7331 \citep[dark gray points,][]{Sutter2022}, resolved star--forming regions from the ``Key Insights in Nearby Galaxies: a Far-Infrared Survey with \textit{Herschel}'' \citep[KINGFISH, light gray triangles,][]{Sutter2019}, global measurements of $z\sim0.02-0.2$ galaxies \citep[brown points,][]{Ibar2015}, and global measurements from local luminous infrared galaxies (LIRGS) from the Great Observatory All--Sky LIRG Survey \citep[GOALS, dark blue points][]{DiazSantos2017}. To match these measurement in a uniform way, we deprojected the infrared surface brightness measurements reported in \citet{Ibar2015} and \citet{DiazSantos2017} by dividing by $\cos{i}$, where $i$ is the galaxy's inclination.  For the sources included in \citet{Ibar2015}, the inclinations were determined by fitting ellipses to the PANSTARRs $r$ band images of each galaxy.  For the galaxies in the GOALS sample, the inclinations were taken from \citet{Kim2013}.  With these updated $\Sigma_{FIR}$ measurements, we see a linear trend between \CII/FIR and $\Sigma_{FIR}$ across three orders of magnitude in $\Sigma_{FIR}$ between our comparison samples. The locus occupied by most of the galaxies of the comparison sample is highlighted in light blue. Of the regions defined in Arp~25, the nucleus and the disk regions fall into the blue locus.
The regions along the shock front and also those immediately after it, the post-shock regions, fall outside the relationship. We also notice that the values of the disk regions, although falling into the blue locus, have a rather high ratio. This is probably due to the high rate of star formation in this galaxy revealed also by optical observations \citep{Tomicic2018}.

We can estimate the excess of [CII] emission by fitting a linear relationship using the disk and nuclear regions, and computing the expected [CII] emission in the other regions based on their surface far-infrared emission. Along the leading edge of the galaxies we find a 60\% excess in [CII] emission, while globally the excess amounts to 25\%. This excess is probably due to the turbulence in the interstellar medium caused by the mechanical dissipation of the shocks due to the impact with the intra-group medium.

\subsection{[CII] and PAH emission}

\begin{figure}
\begin{center}
\includegraphics[width=0.48\textwidth]{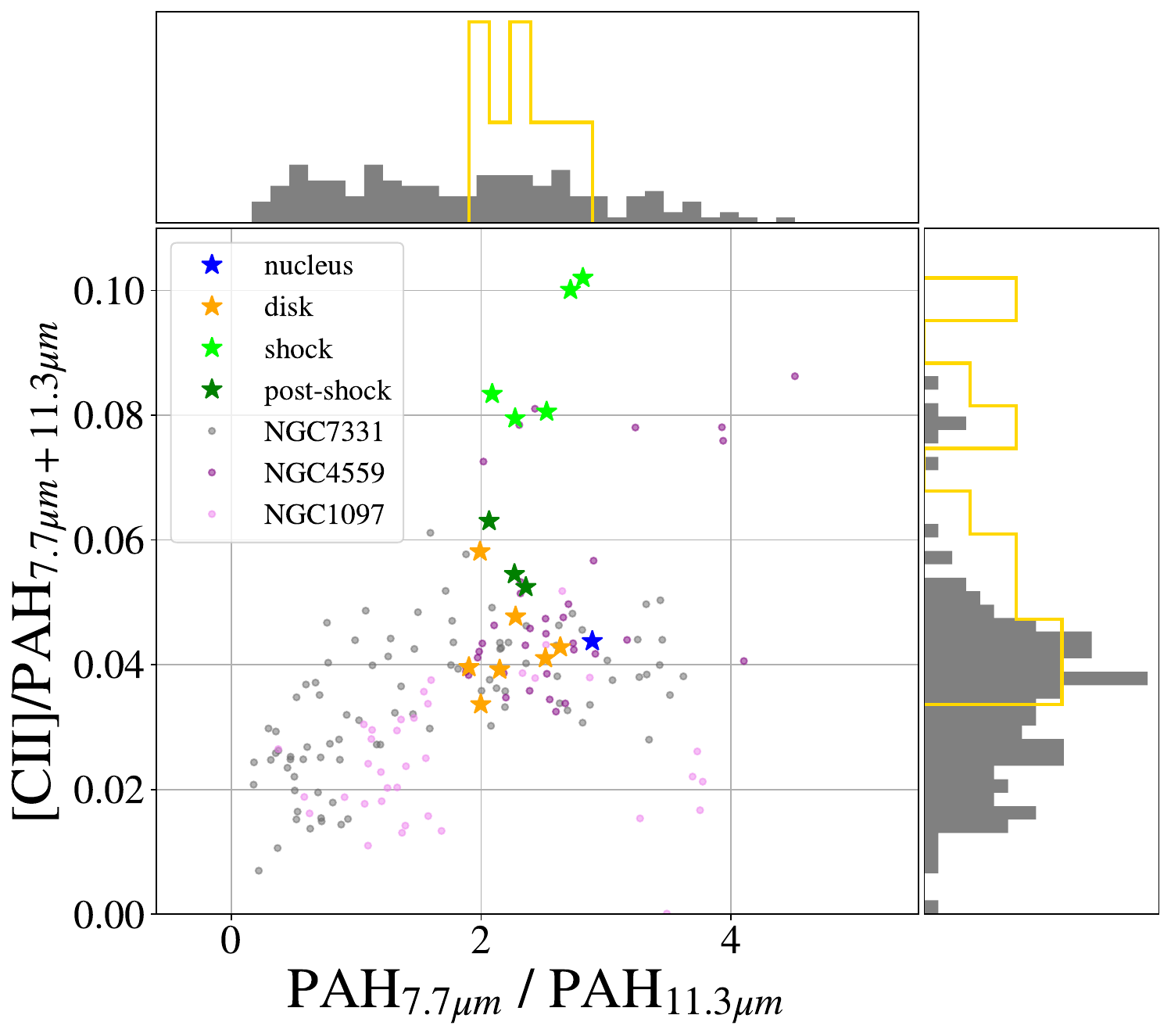}
\end{center}
\caption{The ratio of the 7.7$\mu$m and 11.3$\mu$m PAH feature fluxes versus [CII] to PAH ratio.  The PAH$_{7.7\mu m}$/PAH$_{11.3\mu m}$ ratio is indicative of the average charge of the PAHs.  The histograms on the top and side show the distribution of the comparison samples shaded in gray and the data from Arp 25 outlined in yellow. 
} 
\label{fig:PAHratio}
\end{figure}

In the standard model of [CII] emission, polycyclic aromatic hydrocarbons (PAHs) and dust grains irradiated by the UV light from young stars emit electrons through the photo-electric effect. The collisions of these free electrons with molecules of hydrogen heat the regions of the molecular clouds closest to these stars, called  
photo-dissociation regions (PDRs). Because of the inefficiency of the hydrogen molecule to irradiate energy, due to its lack of dipole, thermal equilibrium is reached thanks to the cooling provided by fine structure lines and predominantly by the singly ionized carbon. Since PAHs provide most of the free electrons in the PDRs, their emission can provide an indicator of the photoelectric efficiency in PDRs more direct than the dust continuum emission. In this scenario, we expect a relationship between PAH and [CII] to be much more stable than that between [CII] and FIR \citep{Croxall2012}. 

Since there are no spectral mid-infrared observations of Arp~25, we used estimates of the 7.7~\micron\ and 11.3~\micron\ PAH features based on the IRAC~4 and WISE~3 band photometry. These estimates are obtained by removing the contributions of stars, large dust grains, and AGN estimated using the SED fits produced by CIGALE. The modeled flux from each of these components is summed, convolved with the transmission function for the specified band, and then subtracted from the observed flux. The remaining emission is then assumed to be only the emission from the 7.7 and 11.3~\micron\ PAH emission features.

Figure~\ref{fig:PAHratio} shows the ratio of the [CII] and the sum of the two PAHs versus the ratio of the 7.7\micron\ and 11.3\micron\ PAH features, an indicator of average PAH charge \citep{Draine2021}. The figure includes data from three other normal star forming galaxies as a comparison: NGC~7331 from \citet{Sutter2022}, NGC~4559 and NGC~1097 from \citet{Croxall2012}.
The figure also includes the histogram of the [CII]/PAH and of the PAH ratio of the comparison sample (grey histogram) and of Arp~25 (yellow histogram).
We can see that the inner and nuclear regions have the same distribution in [CII]/PAH as the bulk of the comparison galaxies (left vertical panel).
The post-shock regions are higher than the rest, but the shock regions have an exceptionally high [CII]/PAH ratio ($\ge 8$\%). It is interesting to note that this difference is not due to a change in the radiation field. In fact, the PAH ratio is substantially the same for the different regions of the Arp~25.
We think this is another evidence of the peculiarity of the regions on the shock front hinting at the non-stellar origin of part of the [CII] emission.

\begin{figure*}
\begin{center}
\includegraphics[width=\textwidth]{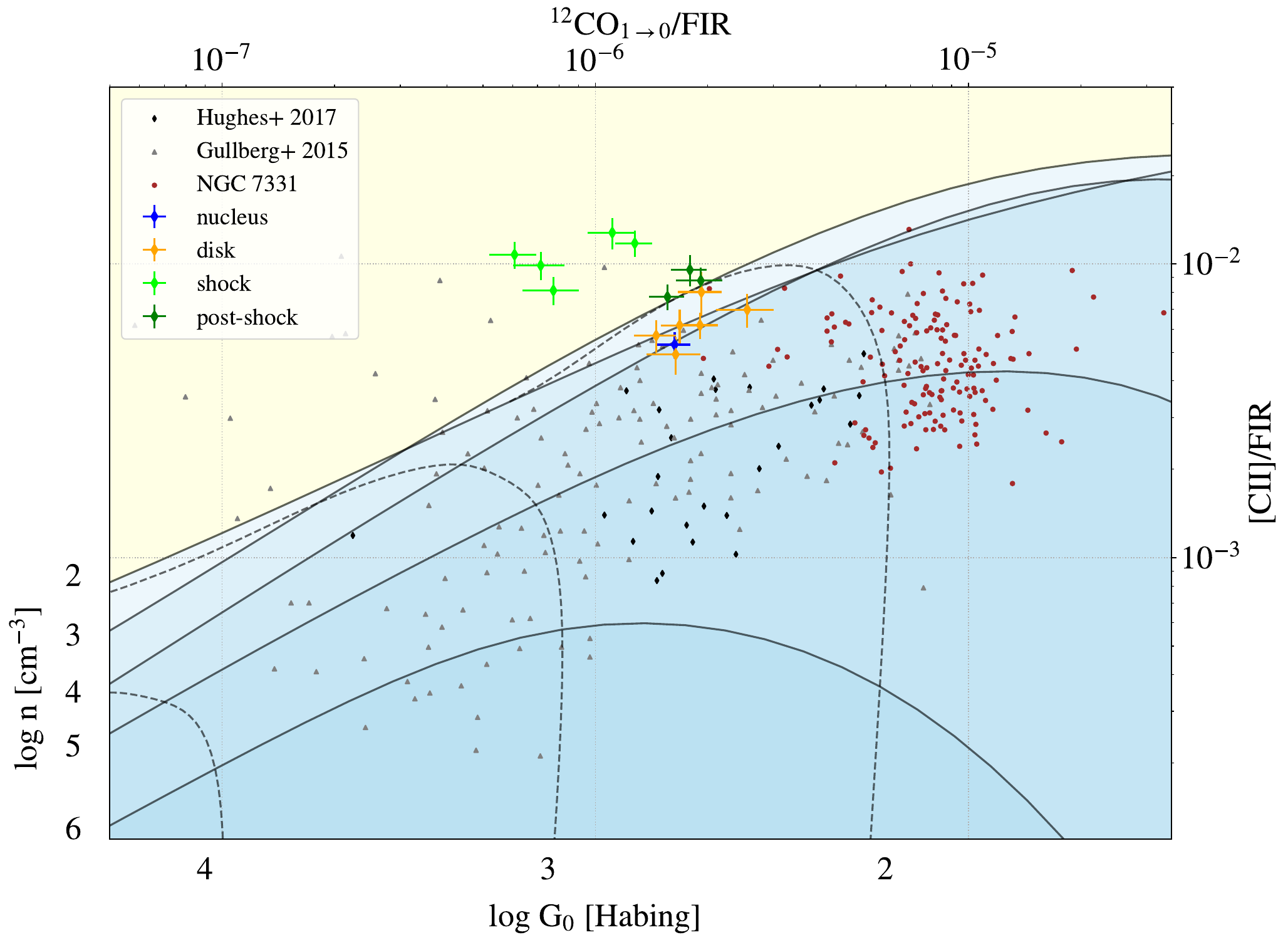}
\end{center}
\caption{The $^{12}$CO$_{1\rightarrow 0}$/FIR values plotted against the [CII]/FIR values of the apertures in Arp~25 over a grid of PDR models of \citet{Kaufman2006} \citep[available in the PDR toolbox, see][]{Pound2022}. The region populated by normal galaxies is shaded in blue. For comparison, we plotted data from regions of the Milky-Way analog NGC~7331 \citep{Sutter2022} and normal star--forming galaxies from \citet{Hughes2017} and \citet{Gullberg2015}. 
}
\label{fig:CIICO}
\end{figure*}

\subsection{[CII] and CO}

Normal star-forming galaxies where the CO and [CII] emission are powered by star formation show a correlation between these two quantities. As shown in Figure~\ref{fig:CIICO}, nearby spiral galaxies included in \citet{Hughes2017} and \citet{Gullberg2015}, as well as regions from the Milky-Way analog NGC~7331 \citep{Sutter2022}, fill a locus in the [CII]/FIR  vs $^{12}$CO$_{1\rightarrow 0}$/FIR plane which can be described with PDR models \citep{Kaufman2006}. A grid of predicted $G_0$, the FUV radiation field in Habing units \citep[typical energy density at the solar circle averaged between 6~eV~$\leq$~h$\nu \leq$~13.6~eV, i.e. 91.2--240~nm, which correspond to $1.6\times 10^{-3}$~erg~cm$^{-2}$~s$^{-1}$, ][]{Habing1968}, and $n$, the gas density in cm$^{-3}$, computed with the PDR toolkit \citep{Pound2022} is shown overlapped to the galaxy values.

Galaxy regions or galaxies falling out of this grid are usually either low-metallicity dwarf galaxies or CO-dark regions \citep{Madden2020}. Another possibility is that the [CII] emission is boosted by an alternative mechanism, such as shocks or turbulence. \citet{Lesaffre2013} showed that even quite low-velocity shocks, passing through a mildly UV-irradiated diffuse (10$^2$--10$^3$ cm$^{-3}$) molecular medium, can produce strong [CII] emission, comparable to other powerful ISM coolants, like mid-IR H$_2$ emission.
Models of this sort were used to explain the powerful H$_2$, [CII] and H$_2$O emission detected by {\it Spitzer} and {\it Herschel} in the shocked filament in Stephan's Quintet \citep{Appleton2017} and in the Hickson compact group 57 \citep{Alatalo2014}.

In order to compare our measurements to PDR models, we multiplied the [CII] fluxes by a factor of 0.75, a typical value of the neutral fraction of [CII] emission, i.e. the part of the emission which is due to non ionized hydrogen \citep[see, e.g., ][]{Sutter2022}.  This allows us to estimate the fraction of the [CII] emission that originates in PDRs.  In addition, we increased the $^{12}$CO$_{1\rightarrow 0}$ fluxes by a factor 2 to account for the likelihood that the $^{12}$CO$_{1\rightarrow 0}$ line will become optically thick in dense star--forming regions \citep{Hughes2017}.  

The emission from the apertures in Arp~25 shows that the radiation field is intense, much more than that of the Milky-Way analog NGC~7331 indicated with red dots. This can account for the high rate of star formation detected with H$\alpha$ images \citep{Tomicic2018}.
But the remarkable result of this comparison is how the apertures along the shock stand clearly out of the region powered by star formation showing that most of the [CII] emission is due to an alternative mechanism.  
Even the regions immediately after the shock are on the border of the relationship showing that the effects of the shocks probably propagates even inside the galaxy, although this can be simply a contamination effect due to the poor spatial resolution of FIFI-LS.

\section{Summary and Conclusions}
\label{sec:conclusions}

We presented new SOFIA observations of the galaxy Arp~25 whose shape is strongly deformed by ram pressure due to its fast motion through the diffuse medium in the NGC~2300 group. We obtained far-infrared images and spectra with the HAWC+ and FIFI-LS instruments. Flux measurements and other quantities derived in the article are reported in Table~\ref{tab:apertures}. We can summarize the main results of this work in the following points:
\begin{itemize}
    \item we gathered a total of 8 galaxies in the NGC~2300 group obtaining a new estimate of the distance of Arp~25 and a virial estimate of the group mass which agrees with previous X-ray studies;
    \item we studied the star formation as a function of the molecular hydrogen mass finding that the star formation is high across the whole galaxy, but it is especially high along the region impacted by the collision with the intra-group medium;
    \item we compared the [CII] emission in different regions of the galaxy to other estimators of photo-electric efficiency such as UV attenuation, dust emission, and PAH emission. We find that the regions along the front of impact with the intra-group medium have a [CII] emission higher than what is expected only from stellar radiation;
    \item the distribution of CO does not show peaks in the impact region as the [CII] intensity does. The comparison of the two emissions against a grid of PDR models shows that the emission from the regions in the impact front cannot be explained with PDR models.
\end{itemize}
We conclude that the impact with the intra-group medium enhances the star formation rate especially along the shock front. However, the enhancement in star formation is not sufficient to explain the high values of [CII] emission detected in the region of the impact. Such a high [CII] emission can be explained as a dissipation of the mechanical energy transferred to the molecular gas by shocks. By assuming a linear relationship between the [CII]/FIR ratio and the FIR surface brightness based on the internal regions of the galaxy, we infer that the [CII] emission is boosted by 60\% along the shock front. This leads to a 25\% increase in the [CII] emission from the whole galaxy.
This observation is the first direct measurement of the enhancement of [CII] emission due to shocks caused by ram pressure in a galaxy group. It clearly shows that the interaction between infalling galaxies and diffuse medium in groups and clusters can significantly alter the total [CII] emission. Since [CII] observations are now routinely used to estimate star formation rates at high redshifts, this study cautions against a direct interpretation of high [CII] fluxes as high star formation rates in clusters of galaxies.

\begin{acknowledgments}
The authors thank S. Shenoy and S. Eftekharzadeh for assistance with the HAWC+ data and the anonymous referee for useful comments and suggestions. This research is based on data and software from: the SOFIA Observatory, operated by USRA (NASA contract NNA17BF53C) and DSI (DLR contract 50OK0901 to the Stuttgart Univ.); the Spitzer Space Telescope, operated by JPL/Caltech under a contract with NASA; WISE, a UCLA--JPL/Caltech project funded by NASA; 2MASS, a NASA/NSF funded project of the Univ. of Massachusetts and IPAC/Caltech; Pan-STARRS1, a survey funded by IfA, Univ. of Hawaii, the Pan-STARRS Project Office, the Max-Planck Society (MPA Heidelberg and MPE Garching), the Johns Hopkins Univ., the Durham Univ., the Univ. of Edinburgh, the Queen's Univ. Belfast, the Harvard-Smithsonian CfA, the LCO Global Tel. Net. Inc., NCU of Taiwan, STScI, NASA grant NNX08AR22G, NSF grant AST-1238877, the Univ. of Maryland, the E\"otv\"os Lor\'and Univ., the Los Alamos Nat. Lab., and the Moore Foundation; GALEX, a NASA small explorer, whose archive is hosted by  HEASARC; the Fabry Perot database at CeSAM/LAM, Marseille, France; the COMING legacy project of the Nobeyama 45m radiotelescope; the WSRT archive  operated by the Netherlands Inst. for Radio Astronomy ASTRON, with support of NWO.
\end{acknowledgments}

%

\facilities{Spitzer (MIPS, IRAC), WISE, Nobeyama, WSRT, Obs. de Haute Provence, GALEX, Pan-STARRS, 2MASS, SOFIA (HAWC+, FIFI-LS)}


\software{astropy \citep{astropy2013,astropy2018}, scipy \citep{2020SciPy}, {\it sospex} \citep[\url{www.github.com/darioflute/sospex}, ][]{Fadda2018}, CIGALE \citep[ \url{https://cigale.lam.fr/}, ][]{Boquien2019}}


 
\begin{deluxetable*}{cccccccccccc}
\label{tab:apertures}
\tabcolsep=0.1cm
\tablewidth{0pt}
\tabletypesize{\footnotesize}
\tablecaption{Arp 25 aperture measurements}
\tablehead{
\colhead{Aper-} &
\colhead{Coordinates} &
\colhead{[CII]$_{158\mu m}$} &
\colhead{$^{12}$CO$_{J=0\rightarrow 1}$}&
\multicolumn{3}{c}{HAWC+}&
\multicolumn{2}{c}{PAH}&
\colhead{FIR}&
\colhead{SFR}&
\colhead{UV att}\\
\colhead{ture}&
\colhead{R.A. -- Dec.}&
\colhead{157.74$\mu$m} &
\colhead{2600.7$\mu$m} &
\colhead{89$\mu$m}&
\colhead{155$\mu$m}&
\colhead{216$\mu$m}&
\colhead{7.7$\mu$m}&
\colhead{11.3$\mu$m}&
\colhead{8-1000$\mu$m}&
\colhead{}&
\colhead{}\\
\colhead{ID} &
\colhead{J2000} &
\colhead{$10^{-16}$W/m$^{2}$}&
\colhead{$10^{-20}$W/m$^{2}$}&
\colhead{Jy}&
\colhead{Jy}&
\colhead{Jy}&
\colhead{10$^8L_\odot$}&
\colhead{10$^8L_\odot$}&
\colhead{10$^8L_\odot$}&
\colhead{M$_{\odot}$/yr}&
\colhead{$10^{40}$erg/s/kpc$^2$}
}
\startdata
1&7:27:12.40 85:45:16.0&9.2$\pm$0.3&10.6$\pm$0.4&2.18&2.67&1.99&3.99&1.38&35.8&0.33&140.3\\
2&7:26:49.69 85:45:53.5&4.5$\pm$0.2&1.2$\pm$0.3&0.38&0.65&0.40&0.83&0.31&9.3&0.15&49.2\\
3&7:26:41.58 85:45:37.2&6.3$\pm$0.2&1.3$\pm$0.3&0.57&0.89&0.70&1.16&0.41&12.0&0.19&64.1\\
4&7:26:40.56 85:45:19.3&4.1$\pm$0.3&1.3$\pm$0.3&0.23&0.76&0.59&0.84&0.40&6.5&0.08&27.7\\
5&7:26:44.71 85:44:56.9&4.8$\pm$0.2&1.7$\pm$0.5&0.53&0.91&0.66&1.08&0.43&12.1&0.16&62.0\\
6&7:27:06.01 85:44:41.6&4.6$\pm$0.1&1.9$\pm$0.2&0.46&0.76&0.59&1.03&0.45&8.1&0.09&34.0\\
7&7:27:00.19 85:44:57.9&5.9$\pm$0.0&4.4$\pm$0.4&0.90&1.51&1.02&1.99&0.85&15.6&0.16&63.4\\
8&7:26:54.39 85:45:14.9&6.0$\pm$0.4&4.2$\pm$0.4&0.57&1.38&0.95&1.63&0.79&12.9&0.14&55.8\\
9&7:26:57.26 85:45:32.8&4.4$\pm$0.2&3.6$\pm$0.7&0.38&1.10&0.82&1.41&0.62&10.2&0.12&43.2\\
10&7:27:22.84 85:45:03.4&4.9$\pm$0.4&4.7$\pm$0.7&1.00&1.55&1.06&2.11&0.80&17.7&0.18&69.9\\
11&7:27:31.84 85:45:17.7&4.2$\pm$0.1&4.9$\pm$0.6&0.74&1.26&0.82&1.87&0.74&14.0&0.15&60.6\\
12&7:27:49.51 85:45:17.5&1.6$\pm$0.2&1.4$\pm$0.3&0.10&0.37&0.39&0.59&0.26&4.1&0.05&19.0\\
13&7:27:13.44 85:45:34.8&2.9$\pm$0.3&3.0$\pm$0.3&0.35&1.03&0.71&1.24&0.65&9.8&0.11&44.5\\
14&7:27:21.28 85:44:46.7&1.4$\pm$0.2&1.8$\pm$0.5&0.28&0.57&0.41&0.71&0.36&5.9&0.07&26.4\\
15&7:27:38.68 85:45:40.1&1.4$\pm$0.1&1.9$\pm$0.6&0.10&0.40&0.29&0.61&0.29&4.1&0.06&17.7\\
16&7:27:54.12 85:45:44.9&1.1$\pm$0.1&&0.06&0.20&0.21&0.33&0.17&2.3&0.03&10.1
\enddata
\tablecomments{Quantities determined using the CIGALE
SED models (PAH$_{7.7}$, PAH$_{11.3}$, FIR, SFR, and UV$_{att}$) have 10\% errors due to the uncertainties introduced during the modelling process. HAWC+ measurements have a 10\% error mainly due to calibration uncertainty.}
\end{deluxetable*}

\bibliography{main}{}
\bibliographystyle{aasjournal}



\end{document}